\documentclass{article}
 
\usepackage{amssymb, a4wide}

\usepackage{epsfig}

\def\be{\begin{equation}}
\def\ee{\end{equation}}
\def\bdm{\begin{displaymath}}
\def\edm{\end{displaymath}}
\def\bea{\begin{eqnarray}}
\def\eea{\end{eqnarray}}
\def\ba{\begin{array}}
\def\ea{\end{array}}

\newcommand{\Bild}[4]{
\begin{figure}[htb]
  \begin{center}
    \leavevmode
    \epsfig{file=#2,height=#1cm}
    \caption{{\small #3}}
    \label{#4}
  \end{center}
\end{figure}}

\begin{document}

\title{\bf An alternative view on quasicrystalline random tilings}

\author{ {\bf Christoph Richard} \\
Institut f\"ur Theoretische Physik, Universit\"at T\"ubingen\\
    Auf der Morgenstelle 14, D-72076 T\"ubingen, Germany}

\maketitle

\begin{abstract}
We apply a framework for the description of random tilings without
height representation, which was proposed recently,
to the special case of quasicrystalline random tilings.
Several important examples are discussed, thereby demonstrating 
the consistency of this alternative description with the conventional one.
We also clarify the latter by deriving a group 
theoretic criterion for the validity of the first 
random tiling hypothesis.
\end{abstract}

\section{Introduction}

The systematic study of random tilings arose from the
insight that they may serve as models for entropically
stabilized quasicrystals \cite{Els85a, Els85b}.
As such, they are an important and interesting alternative to the
perfect tilings based upon the projection method.
Although random tilings were used intensively to study
transport properties and dynamical properties of real
quasicrystals, not much effort was spent to deduce
random tiling properties from general grounds \cite{Hen88, Hen91}.
We mention the two random tiling hypotheses stated
by Henley in 1991, which serve as a starting point to
infer diffraction properties, using ideas of
the Landau theory of phase transitions.
Recently, a more general approach was investigated
\cite{RHHB98, Ric99, Hof97}, mainly for two reasons:
On the one hand, the concept of entropic stabilization applies
also to crystalline solids which therefore should be included
in a more general description.
On the other hand, the quasicrystalline random tilings
share a special feature which stems from the fact that
these tilings are derived from the perfect ones:
They have a height representation -- each tiling can
be embedded as a surface in a higherdimensional space.
Since tiling surfaces with equal (mean) density of prototiles
have equal (mean) slope, the slope parameters can be used
to describe the random tiling ensemble.
This is the so-called {\it phason strain} parametrization
which governs the conventional description.

This description has a number of shortcomings:
First of all, it only allows the description of those random 
tilings which do allow a height representation -- being the
vanishing (though interesting) minority of all possible tilings.
Second, symmetry analysis in this framework is constrained
to {\it geometric} symmetries, whereas there may be other relevant
symmetries of the tiling ensemble.
An example are colour symmetries modelling chemical disorder,
which is shown to play an important role for real quasicrystals 
\cite{JB96, JRB97}. 
Third, it was expected that a more general analysis could also
clarify the origin of the random tiling hypotheses mentioned above.

This was the viewpoint which led to the random tiling description
proposed in \cite{RHHB98}.
There, the grand-canonical tiling ensemble was considered where
prototiles are energetically degenerate, and the chemical potentials
of the different prototiles or their (mean) densities are the
only macroscopic parameters.
It was explained how to perform a symmetry analysis in this
framework.
This led to a proof of a generalization of the first random
tiling hypothesis which asserts that the point of maximum entropy
is a point of maximum symmetry.
In addition, the range of validity of the second hypothesis could
be analyzed using the grand-canonical setup, and an exactly
solved (crystallographic) counterexample could be given.
A rigorous treatment of diffraction is also possible using this setup 
\cite{BH99}.

The examples presented so far were taken only from the crystallographic case.
It remained to apply the proposed description scheme to
quasicrystallographic random tilings and to show the consistency
of the two approaches.
This is where this paper aims at:
The next two chapters are devoted to the description of two important 
examples of quasicrystallogrphic random tilings, the Ammann-Beenker tiling 
\cite{LPW92} and the square-triangle tiling \cite{OH93, Wid93, Kal94, Kal97}.
We perform a symmetry analysis along the lines of \cite{RHHB98} and
compare the approach with the conventional one, thereby showing the
consistency of the two approaches.
In the Appendix, we shortly review and clarify the conventional 
description.
We analyze the connection between maximal symmetry and vanishing phason 
strain and derive a group theoretic criterion.
If the criterion is valid -- which is the case in all examples
we met -- the first random tiling hypothesis is fulfilled in its
original formulation:
The point of maximum entropy is a point of maximum symmetry \cite{RHHB98},
and maximum symmetry in turn implies vanishing phason strain.  
A comparing discussion of the two approaches concludes the paper.

\section{The Ammann-Beenker Random Tiling}

The prototiles of the Ammann-Beenker random tiling are
two squares and four $45^\circ$-rhombi as shown in Fig.~\ref{fig:ammtiles}.
There is no further matching rule apart from the face-to-face tiling
condition, which imposes the relation $\sum \rho_i + \sum \sigma_i =1$
on the prototile densities.
\Bild{3}{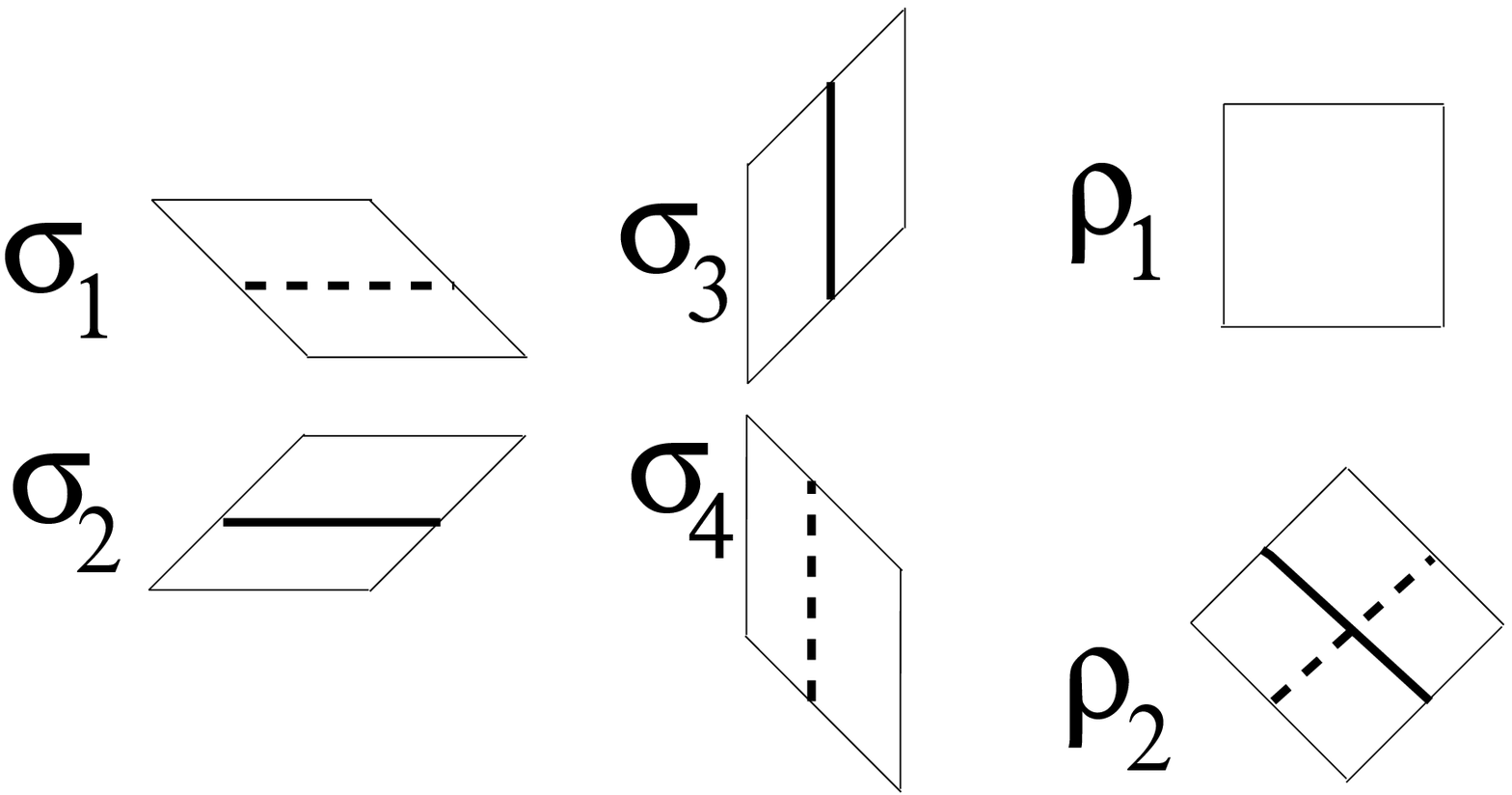}{Prototiles of the Ammann-Beenker random tiling}{fig:ammtiles}
Tilings can alternatively be viewed as world lines of
particles of two different kinds, as is indicated by the
decoration.
It was shown in \cite{LPW92} that there is a nonlinear
constraint among the different prototile densities
which reduces the number of independent variables to four.
This constraint was obtained using the height representation 
of this model, which is given below.
It should be noted that it can be independently derived relating 
the tile description to the line interpretation.
This method was introduced in \cite{GN97}.
The constraint reads
\be
\rho_1 \rho_2 = 2 \left( \sigma_1\sigma_3+\sigma_2\sigma_4 \right).
\ee
The derivation imposes periodic boundary conditions.
It is highly plausible that the constraint holds (asymptotically)
also in the case of free boundary conditions, though there is no
proof of this assumption.
Complementary, the line representation suggests four independent
parameters since it is possible to fix the densities of different 
lines as well as their mean direction independently.
Moreover, care has to be taken in choosing a set of independent
parameters:
The four rhombi densities, for example, allow the determination
of the square densities only up to a permutation, due to the
nonlinear constraint.
This subtlety does not arise if, for example, three rhombi densities
and one square density are chosen.
The quadratic invariants given below are insensitive to this
because only absolute values of square density differences occur.

At the point of maximum symmetry, the four rhombi densities
and the two square densities have the same value.
The nonlinear constraint fixes them to
\be 
\rho_i=1/4, \qquad \sigma_i=1/8.
\ee
This is the point where the squares and the rhombi each occupy
half of the tiled area.
We determine the second order expansion of the entropy density 
according to ${\cal D}_8$-symmetry.
We introduce reduced prototile densities
\be
r_i=\rho_i-1/4, \qquad s_i=\sigma_i-1/8,
\ee
such that the point of maximum symmetry is the origin.
The vector space relevant for the symmetry analysis is obtained
through {\it linearization} of the constraints.
It is four-dimensional and determined by
\be
r_1+r_2=0, \qquad s_1+s_2+s_3+s_4=0.
\ee
The analysis of the symmetries in the vector space of the independent,
reduced prototile densities leads to a second order entropy expansion
of the form
\be
\label{form:amment}
s(r,s) = s_0 - \frac{1}{2}\sum_i\lambda_i I^{(i)}(r,s) + \ldots
\ee 
There are three (positive) elastic constants $\lambda_i$ with invariants
\bea
I^{(1)}(r,s) &=& 4 \, r_1^2, \\
I^{(2)}(r,s) &=& (s_1+s_3)^2, \nonumber \\
I^{(3)}(r,s) &=& \frac{1}{2}(s_1-s_3)^2 + (s_1+2s_2+s_3)^2. \nonumber
\eea
This can be obtained by decomposing the representation of the
symmetry group in the vector space of the reduced, independent
density parameters according to \cite{RHHB98}, in full analogy
to the height representation treatment which is presented in
the Appendix. 

We now derive the relation to the elastic constants defined
from the height representation approach.
This is done by expressing the expansion (\ref{form:amment})
in terms of the slope parameters of the height representation.
The positions of tiling vertices are integer linear combinations of
vectors of the regular 8-star.
They therefore belong, viewed as complex numbers, to the 
module ${\mathbb Z}[\xi]$ with $\xi=e^{2\pi i/8}$.
We characterize the height function algebraically.
To this end, we recall some facts concerning the associated
cyclotomic field \cite{LP84}.
$\xi$ is a root of the eighth cyclotomic 
polynomial $P_{8}(x)=x^4+1$.
The other roots are $\xi^3$, $\xi^5=-\xi$ and $\xi^7=\bar{\xi}$.
The automorphism group on this set of primitive roots of unity is the
Galois group ${\cal G} \simeq {\cal C}_2\times {\cal C}_2$.
The automorphisms can be uniquely extended to automorphisms of the corresponding 
module.
The usual height function $h$ on the set of possible vertex positions
is defined by
\be
h(\alpha) = \sigma(\alpha)
\ee
with $\alpha \in {\mathbb Z}[\xi]$, $ {\cal G} \ni \sigma : \xi \mapsto \xi^5$.
The set of all embedded vertex positions constitutes the four-dimensional
primitive hypercubic lattice ${\mathbb Z}^4$, and each tiling corresponds 
to a two-dimensional surface.
This is the geometric origin of the height representation.
It is possible to define a height function on the prototiles by
linear extrapolation of the height function on the corresponding vertices.
The new macroscopic parameters to describe the tiling ensemble
are the components of the {\it phason strain tensor}
$$E = \left(\begin{array}{cc} E_{1,1} & E_{1,2} \\ E_{2,1} & E_{2,2} \end{array}\right)$$
in cartesian coordinates.
The phason strain is defined as (mean) slope of the height function.
This leads to the relations
\bea
\label{form:ammrel}
E_{1,1} &=& \rho_1-\rho_2+(\sigma_1+\sigma_2)-(\sigma_3+\sigma_4), \\
E_{2,2} &=& \rho_1-\rho_2-(\sigma_1+\sigma_2)+(\sigma_3+\sigma_4), \nonumber\\
E_{1,2} &=& 2(\sigma_4-\sigma_3), \nonumber\\
E_{2,1} &=& 2(\sigma_1-\sigma_2). \nonumber
\eea
We observe $E=0$ at the point of maximal symmetry, which already follows
from group theoretical considerations given in the Appendix.
In order to write the entropy expansion (\ref{form:amment}) in terms of the
phason strain, the relation between phason strain and independent
tile densities has to be invertible at the point of maximum symmetry.
This is not the case if the four rhombi are taken as parameters.
If three rhombi densities and one square densities are taken instead,
the relation (\ref{form:ammrel}) is invertible in the whole phase space.
This leads to an expansion
\bea
s(E) &=& s_0 -\frac{1}{2} K_1 (E_{1,1}+E_{2,2})^2 
 -\frac{1}{2} K_2 (E_{1,2}-E_{2,1})^2 \\
 &&-\frac{1}{2} K_3 \left( (E_{1,1}-E_{2,2})^2 + (E_{1,2}+E_{2,1})^2 \right) + \ldots
\nonumber
\eea
This expression is known from the literature \cite{LPW92} and is rederived
in the Appendix using the symmetry analysis in the height representation.
The relation between the elastic constants is
\be
\lambda_1 = 4 K_1 > 0, \qquad \lambda_2= 16 K_2 > 0, \qquad \lambda_3 = 16 K_3 > 0.
\ee

\section{The Square-Triangle Random Tiling}

The prototiles of the square-triangle random tiling are
three squares and four equilateral triangles as shown in Fig.~\ref{fig:sqtiles}.
\Bild{2}{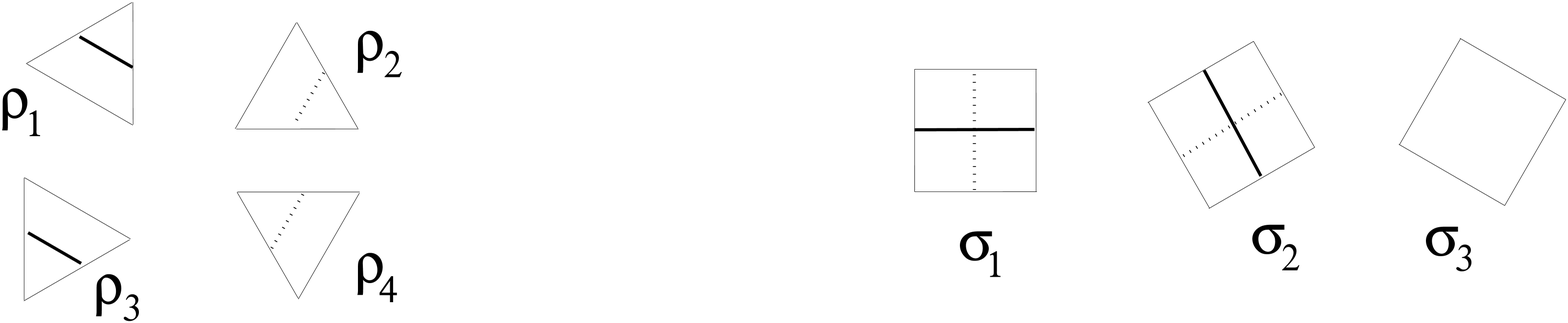}{Prototiles of the square-triangle random
tiling}{fig:sqtiles}
There is no further matching rule apart from the face-to-face tiling
condition, which imposes the relation $\sum \rho_i + \sum \sigma_i =1$
on the prototile densities.
Tilings can alternatively be viewed as world lines of
particles of two different kinds, as is indicated by the
decoration.
Using this representation, it is easy to see that triangles
always occur in pairs.
Between the (mean) prototile densities we therefore have 
the relations $\rho_1=\rho_3$ and $\rho_2=\rho_4$.
It was shown in \cite{Gie98} that there is a nonlinear
constraint among the different prototile densities
which reduces the number of independent variables to three.
The constraint reads
\be
16 \, \rho_1 \rho_2 = 3 \left( \sigma_1\sigma_2+\sigma_2\sigma_3+\sigma_3 \sigma_1 \right).
\ee
It should be mentioned that the derivation imposes periodic
boundary conditions.
The same constraint is believed to hold (asymptotically)
in the case of free boundary conditions.
This is suggested from the complementary observation that
the line representation allows three independent parameters:
the densities of the lines of different type and the frequency
of one type of crossing.
Due to the nonlinear constraint, care has to be taken in choosing 
a set of independent parameters:
The three square densities, for example, allow the determination
of the triangle densities only up to a permutation.
This subtlety does not arise if two square densities
and one triangle density are chosen instead.
The quadratic invariants given below are insensitive to this
because only absolute values of triangle density differences occur.

At the point of maximum symmetry, the four triangle densities
and the three square densities have the same value.
The nonlinear constraint fixes their values to
\be 
\rho_i=1/8, \qquad \sigma_i=1/6.
\ee
This corresponds to the situation where the squares (resp. triangles)
occupy half of the tiled area.
We determine the second order expansion of the entropy according to
${\cal D}_{12}$-symmetry.
We introduce reduced prototile densities
\be
r_i=\rho_i-1/8, \qquad s_i=\sigma_i-1/6,
\ee
such that the point of maximum symmetry is the origin.
The vector space relevant for the symmetry analysis is obtained
through {\it linearization} of the constraints.
It is three-dimensional and determined by
\be
r_1+r_2=0, \qquad s_1+s_2+s_3=0.
\ee
The analysis of the symmetries in the vector space of the independent,
reduced prototile densities leads to an entropy expansion
of the following form, up to second order:
\be
\label{form:sqtrent}
s(r,s) = s_0 - \frac{1}{2}\sum_i\lambda_i I^{(i)}(r,s) + \ldots
\ee 
There are two (positive) elastic constants $\lambda_i$ with invariants
\bea
I^{(r)}(r,s) &=& 4r_1^2, \\
I^{(s)}(r,s) &=& 6(s_1^2 - s_1 s_2 + s_2^2). \nonumber
\eea
The invariants are obtained from the decomposition
of the representation of the symmetry group into its irreducible
components, according to \cite{RHHB98}.
We now derive the relation to the elastic constants of the height 
representation approach.
The positions of tiling vertices are integer linear combinations of
vectors of the regular 12-star.
They therefore belong, viewed as complex numbers, to the 
module ${\mathbb Z}[\xi]$ with $\xi=e^{2\pi i/12}$.
We characterize the height function algebraically.
Therefore, we remind of some facts concerning the associated
cyclotomic field \cite{LP84}.
$\xi$ is a primitive root of unity, annihilating the 12th cyclotomic 
polynomial $P_{12}(x)=x^4-x^2+1$.
The other roots are $\xi^5$, $\xi^7=-\xi$ and $\xi^{11}=\bar{\xi}$.
The automorphism group on this set of primitive roots of unity is the 
Galois group ${\cal G} \simeq {\cal C}_2\times {\cal C}_2$.
The automorphisms can be uniquely extended to automorphisms of the corresponding 
module.
The usual height function $h$ on the set of possible vertex positions
is defined by
\be
h(\alpha) = \sigma(\alpha)
\ee
with $\alpha \in {\mathbb Z}[\xi]$, $ {\cal G} \ni \sigma : \xi \mapsto \xi^7$.
It is possible to define a height function on the prototiles by
linear extrapolation of the height function on the corresponding vertices.
The new macroscopic parameters to describe the tiling ensemble
are the components of the {\it phason strain tensor}
$$E = \left(\begin{array}{cc} E_{1,1} & E_{1,2} \\ E_{2,1} & E_{2,2} \end{array}\right)$$
in cartesian coordinates.
The phason strain is defined as (mean) slope of the height function.
This leads to the relations
\bea
\label{form:sqtrrel}
E_{1,1} &=& \sigma_1 - \frac{1}{2}(\sigma_2+\sigma_3)+
(\rho_1+\rho_3)-(\rho_2+\rho_4), \\
E_{2,2} &=& -\sigma_1 + \frac{1}{2}(\sigma_2+\sigma_3)+
(\rho_1+\rho_3)-(\rho_2+\rho_4), \nonumber \\
E_{1,2} &=& E_{2,1}= \frac{\sqrt{3}}{2}(\sigma_3-\sigma_2). \nonumber
\eea
We find $E=0$ at the point of maximal symmetry, which already follows
from group theoretical considerations given in the Appendix.
Since the model has three degrees of freedom, as follows from the density
analysis, we have to expect a special property of the height function
which reduces the number of independent slope parameters,
in contrast to the previous example.
It is indeed easily seen that each height function has the {\it irrotationality} 
property \cite{OH93}
\be 
\Re \left(\oint h(z)\, d\bar{z} \right) = 0. \label{form:irro}
\ee
The equality of the off-diagonal elements is a consequence
of the irrotationality property (\ref{form:irro}).
(It is also possible to define a height function by
$ {\cal G} \ni \sigma : \xi \mapsto \xi^5$.
In this case, the irrotationality property reads 
$\Re \left(\oint h(z)\, d z \right) = 0$.)

In order to write the entropy expansion (\ref{form:sqtrent}) in terms of the
phason strain, the relation between phason strain and independent
tile densities has to be invertible, at least at the point of maximum symmetry.
If the three squares are taken as parameters, this is not the case in the phase 
with ${\cal C}_4$-symmetry.
If two square densities and one triangle densities are taken instead,
the relation (\ref{form:sqtrrel}) is invertible in the whole phase space.
This leads to an expansion
\be
s(E) = s_0 -\frac{1}{2} K_{\mu} \left( {\rm tr}(E)\right)^2 + 
\frac{1}{2} K_{\xi} \det(E) + \ldots,
\ee
which is known from the literature \cite{OH93} and is rederived
in the Appendix using the symmetry analysis in the height representation.
The relation between the elastic constants is
\be
\lambda_s = \frac{K_{\xi}}{2} > 0, \qquad \lambda_r=4(4 K_{\mu}-K_{\xi})>0.
\ee

\section{Conclusion}

We performed a symmetry analysis of well-known planar quasicrystalline 
random tilings using the prototile density approach.
We were able to show the consistency to the conventional
approach via the height representation of these models.
For the square-triangle tiling, the approach
illuminated the irrotationality property whose
appearance was rather unexpected before.
Whereas it seems to be natural to describe random tilings
using the prototile densities as macroscopic parameters,
the approach is commonly hard to follow because it requires
the knowledge of all (possibly nonlinear) constraints in order
not to overestimate the number of elastic constants.
The height representation, in turn, has the disadvantage
that is is not obvious whether the slope parameters 
are independent and whether they constitute a sufficient 
set to describe all degrees of freedom of the ensemble.
Furthermore, to describe other than geometric symmetries,
one has to derive the relation between phason strain and
the prototile densities explicitly.
Therefore, the two approaches are in fact complementary
in deriving properties of random tilings.
The exactly solved eightfold random tiling, for example,
must possess a constraint additional to the ones derived
in \cite{GN97}, since the height representation
and the density description yield different numbers of
elastic constants otherwise, as can be shown.
Research in this direction is in progress.

\section*{Acknowledgments}

The author is grateful for discussions with Michael Baake,
Jan de Gier, Joachim Hermisson and Moritz H\"offe.
He also thanks Michael Baake for several useful
comments on the manuscript.
Financial support from the German Science Foundation (DFG)
is gratefully acknowledged.

\section*{Appendix: Symmetry and Phason Strain}

We shortly expose the symmetry analysis of random tilings which
permit a height representation, in the framework of our viewpoint
exposed elsewhere \cite{RHHB98}.
In particular, this will lead to a criterion for the validity of 
the first random tiling hypothesis \cite{Hen91} which states that
the point of maximum entropy occurs at vanishing phason strain.

Let a random tiling ensemble of the $d_{||}$-dimensional vector 
space $V_{||}$ be given.
A height representation assigns to each tiling a $d_{||}$-dimensional
surface in a higher-dimensional vector space $V=V_{\|} \oplus V_{\bot}$
with the property that surfaces of tilings with equal prototile 
(mean) density have the same (mean) slope in $V$.
In this situation, the (mean) surface slope can be used to parametrize
the grand-canonical tiling ensemble.
We discuss the influence of geometric symmetries on this description.
The (mean) tiling slope can be represented by a linear map 
$E: V_{\|} \to V_{\bot}$, usually called {\it phason strain}.
Symmetries are bijections on the (mean) prototile densities which leave
the entropy density invariant.
Let $D_\|$ denote the representation of the geometric symmetries
in the physical space $V_\|$.
The height representation induces a representation $D_\perp$ of the geometric
symmetries in the internal space $V_\perp$.
The phason strain tensor transforms under geometric symmetries according to
\be
{\tilde E} = D_\bot E D_{||}^{-1}.
\ee
We now focus on the point of maximum symmetry.
$E$ is invariant at this point, since this is by definition true of the 
(mean) prototile densities, which determine $E$ uniquely.
If the representations $D_{||}$ and $D_\bot$ are irreducible and not
equivalent, Schur's lemma yields $E=0$.
In the general case we have
\begin{itemize}
\item The phason strain tensor vanishes at maximal symmetry if no irrep
in internal space is equivalent to an irrep in physical space.
\end{itemize}
If this criterion is fulfilled, the first random tiling hypothesis
in the formulation of Henley is satisfied since the point of maximum entropy 
is always a point of maximum symmetry, as was shown in \cite{RHHB98}.
On the other hand, the criterion is fulfilled in all situations we
met due to the special construction of the height function.

Symmetry analysis may alternatively be performed in the tensor product
$V_{\|}\otimes V_{\bot}$, which is most advantageous when discussing
the influence of symmetries on the entropy density.
In this case, $E$ is regarded as a vector with components of the
matrix $E:  V_{\|} \to V_{\bot}$.
The geometric symmetries are represented via
\be
{\tilde E} = D \cdot E = \left( \left( D_{\|}^{-1}\right)^{t} \otimes D_{\bot} \right) \cdot E.
\ee
Since the point of maximum symmetry is by definition a fixed point,
we conclude:
\begin{itemize}
\item The phason strain at the point of maximum symmetry consists of
components in direction to the trivial 1d-irreps of the representation
in $V_{\|}\otimes V_{\bot}$.
\item In particular, the phason strain vanishes at maximal symmetry,
if the representation in $V_{\|}\otimes V_{\bot}$ does not contain
trivial parts.
\end{itemize}
This criterion is equivalent to the statement given above,
as follows from a closer look at the Clebsch-Gordan decomposition
of the product representation.

We give an expansion of the entropy density of the random tiling ensemble
in terms of the phason strain at $E=0$.
Without loss of generality we assume that the symmetries are represented by 
orthogonal mappings in $V_{\|}\otimes V_{\bot}$.
The entropy density is invariant under symmetry transformations,
$s({\tilde E}) = s(E)$.
Since $E=0$ is a fixed point of the action of the symmetry group,
the symmetries commute with the Hessian $H$, the second derivative
of the entropy density,
\be
\left. H \right|_{E=0} D = D \left. H \right|_{E=0}.
\ee
According to Schur's lemma, $\left. H \right|_{E=0}$ acts trivially
on the irreducible components in $V_{\|}\otimes V_{\bot}$.
Therefore,  $\left. H \right|_{E=0}$ is a linear combination of
projectors $P^{(i)}$ onto the irreducible components,
\be
\left. H \right|_{E=0} = - \sum_i \lambda_i P^{(i)}.
\ee
The (positive) $\lambda_i$ are the elastic constants of
the random tiling, arising from the height representation.
The second order term of the entropy expansion is of the form
\be
s_2(E,E) = -\frac{1}{2} \sum_i \lambda_i I^{(i)}(E,E),
\ee
where $I^{(i)}(E,E)$ denotes the quadratic invariants
\be
I^{(i)}(E,E) = \langle E, P^{(i)} E\rangle = ||P^{(i)}E ||^2.
\ee
$\langle \cdot, \cdot \rangle$ denotes the scalar product in
$V_{\|}\otimes V_{\bot}$ and $||\cdot ||$ the induced norm.

We treat as example two-dimensional random tilings with dihedral 
symmetry ${\cal D}_n$.
This comprises all models discussed above.
In the physical space, generators of the symmetries may be
given by a rotation $R_{\|}$ and a reflection $S_{\|}$.
We assume that the height representation induces a representation
in the internal space with generators $R_{\bot}=-R_{\|}$
and $S_{\bot}=S_{\|}$.
(This is the canonical choice for $n$ even.)
Since the corresponding representations $D_{\|}$ and $D_{\bot}$ 
are irreducible and not equivalent, we immediately conclude $E=0$ 
at maximal symmetry.
Let us determine the quadratic invariants.
The representation of the symmetry group in $V_{\|}\otimes V_{\bot}$
decomposes into three parts, a two-dimensional rotation and two
one-dimensional reflections.
If we introduce standard bases, this yields three elastic constants 
with invariants \cite{Hen91}
\bea
I^{(1)}(E,E) &=& \frac{1}{2} (E_{1,1} + E_{2,2})^2, \\
I^{(2)}(E,E) &=& \frac{1}{2} (E_{1,2} - E_{2,1})^2, \nonumber\\
I^{(3)}(E,E) &=& \frac{1}{2} (E_{1,1} - E_{2,2})^2 + \frac{1}{2} (E_{1,2} + E_{2,1})^2.
\nonumber
\eea
Mind that the square-triangle tiling is described by only two
elastic constants, since the off-diagonal elements in $E$ are
equal, due to the irrotationality property (\ref{form:irro}).

\end{document}